\documentclass[11pt]{article}
\usepackage[margin=1in]{geometry}
\usepackage{amsmath,amssymb}
\usepackage{algorithm} \usepackage{algpseudocode} 
\usepackage{graphicx}
\usepackage{pdflscape}
\usepackage{booktabs}
\usepackage{multirow}
\usepackage{xspace}
\usepackage[super,comma,sort&compress]{natbib}
\usepackage{xcolor}
\usepackage[table]{xcolor}
\usepackage{bm}

\usepackage[hidelinks]{hyperref}
\usepackage{xurl}
\usepackage{tikz}
\usetikzlibrary{arrows.meta}
\usepackage[T1]{fontenc}
\usepackage{lmodern}
\usepackage{placeins}
\usepackage{float}
\definecolor{gen}{HTML}{22648C}
\definecolor{load}{HTML}{A76320}
\definecolor{opened}{HTML}{B43136}
\definecolor{opened}{HTML}{B43136}

\tikzset{
 bus/.style={circle,draw,thick,minimum size=9mm,font=\bfseries},
 generator/.style={bus,draw=gen},
 demand/.style={bus,draw=load},
 line/.style={-{Stealth[length=2.5mm]},thick},
 binding/.style={line,very thick,draw=red},
 flow/.style={fill=white,inner sep=3pt,font=\small},
 offline/.style={draw=opened,dashed,thick},
 info/.style={align=center,font=\small}
}
\newcommand{\networkdraw}[7]{%
\begin{minipage}[t]{0.325\textwidth}
\centering
{\large\bfseries Topology #1}\par\smallskip
#2\par\vspace{6mm}
\resizebox{\linewidth}{!}{%
\begin{tikzpicture}[x=1cm,y=1cm]
\path[use as bounding box] (-1,-1.2) rectangle (4.8,5.3);
\node[generator] (b1) at (0,3.4) {1};
\node[generator] (b2) at (3.8,3.4) {2};
\node[demand] (b3) at (0,0) {3};
\node[demand] (b4) at (3.8,0) {4};
\node[info,above=4mm] at ([xshift=4mm]b1.north) {cheap gen\\\$10/MWh\\$g_1=#3$ MW};
\node[info,above=4mm] at ([xshift=-4mm]b2.north) {expensive gen\\\$100/MWh\\$g_2=#4$ MW};
\node[info,below=4mm] at (b3.south) {load #5 MW};
\node[info,below=4mm] at (b4.south) {load #6 MW};
\draw[binding] (b1) -- node[flow,above] {$30/30$} (b2);
\ifcase#1\relax
\or
 \draw[line] (b1) -- node[flow,rotate=90,above] {$44/50$} (b3);
 \draw[line] (b2) -- node[flow,sloped,above] {$14/60$} (b3);
 \draw[line] (b2) -- node[flow,rotate=90,below] {$42/90$} (b4);
 \draw[line] (b3) -- node[flow,above] {$28/80$} (b4);
\or
 \draw[binding] (b1) -- node[flow,rotate=90,above] {$50/50$} (b3);
 \draw[line] (b2) -- node[flow,rotate=90,below] {$50/90$} (b4);
 \draw[line] (b3) -- node[flow,above] {$30/80$} (b4);
\or
 \draw[binding] (b1) -- node[flow,rotate=90,above] {$50/50$} (b3);
 \draw[line] (b2) -- node[flow,sloped,above] {$20/60$} (b3);
 \draw[line] (b2) -- node[flow,rotate=90,below] {$30/90$} (b4);
\fi
\end{tikzpicture}%
}
\par\medskip
{\large $\text{cost}=\$#7$}\par
\end{minipage}%
}

\newcommand{\model}{\textsc{Line-switching Ranking Proxy}\xspace}

\title{Machine learning-guided real-time AC optimal transmission switching}
\author{%
Weimin Huang\textsuperscript{1,2,*}\quad
Weiwei Yang\textsuperscript{1,3,*,\dag}\quad
Bistra Dilkina\textsuperscript{2}\quad
Konstantina Mellou\textsuperscript{1}\\
Ishai Menache\textsuperscript{1}\quad
Marco Molinaro\textsuperscript{1}\quad
Chris White\textsuperscript{1}\quad
Baosen Zhang\textsuperscript{1,4}\\[6pt]
\small\textsuperscript{1}Microsoft Research\quad
\textsuperscript{2}University of Southern California\quad
\textsuperscript{3}asemic.ai\quad
\textsuperscript{4}University of Washington\\[2pt]
\small\textsuperscript{*}Equal contribution, listed alphabetically.\\[2pt]
\small\textsuperscript{\dag}Corresponding author: \href{mailto:weiwei@asemic.ai}{weiwei@asemic.ai}}

\date{}

\begin{document}
\maketitle

\begin{abstract}
\noindent

The operation of power grids has substantial economic consequences, affecting generation costs and, ultimately, electricity prices. The Optimal Transmission Switching (OTS) problem seeks to reduce these costs by selectively switching off transmission lines. However, OTS gives rise to a challenging mixed-integer nonlinear optimization problem that is difficult to solve at operationally relevant timescales, particularly as rapidly changing loads create a need for real-time switching decisions. To address this challenge, we develop a machine learning-guided method that identifies high-quality line switching decisions in real time. Specifically, we construct a grid-specific switching-value model that rapidly estimates the generation-cost reduction associated with opening transmission lines by fine-tuning a pretrained model for AC optimal power flow (AC-OPF). The switching-value model provides signals to guide promising switching decisions, and topologies proposed by the model are evaluated using exact AC-OPF solvers to verify feasibility and obtain the actual generation cost. We develop multiple strategies that use the ML surrogate to efficiently explore the combinatorial space of transmission topologies, including Fixed-Rank Search, Adaptive-Rank Search, and Swap-Enhanced Search. Across multiple benchmark networks, our proposed algorithms identify line switching decisions that significantly reduce generation cost within a 50s computational budget. In particular, under a realistic data center build-out scenario on the 2,000-bus synthetic Texas grid, our approach reduces generation costs by 31.97\% and gross nodal load payments by 94.34\%. More broadly, our results demonstrate that a fine-tuned surrogate can effectively guide search even without guaranteeing feasibility, enabling classical optimization methods to tackle large nonlinear combinatorial problems on operationally relevant timescales.

\end{abstract}

\bigskip

\noindent
\section{Introduction}

The cost and reliability of the electric grid have taken on a renewed importance around the world. In developed economies, electricity prices are now a central cost-of-living issue~\cite{oshaughnessy2026demandgrowth,tausendfreund2025european,reuters2025pjmprices}. In other regions, the ability to provide access to electric power is fundamental for health and economic development~\cite{iea2026electricity,lukuyu2023powering}. At the same time, the cost of electricity has increased at a rate that is much faster than inflation~\cite{wiser2025electricityprices,acer2026energyprices}. 

There are many intertwined causes of the increase in electricity costs, but one of the main drivers is the dramatic increase in load. For example, in the US, approximately 10 GW of data center capacity entered the system in 2025~\cite{FERC2026StateOfMarkets}. 
Interestingly, generation in the US has kept up with load, with a total of 13.7 GW net additional capacity deployed in 2025, most of which is distributed solar and storage~\cite{Peterson2026USGeneration}. The part of the grid that has not changed is the transmission infrastructure. In 2025, the continental US constructed less than 900 miles of high voltage transmission lines~\cite{EIAOpenData}. To put this number in perspective, it is less than 0.3\% of the US's transmission infrastructure~\cite{DOE2017QER}. Therefore, there is a severe misalignment between rapid changes in load and generation and stagnation in transmission. 

Addressing the root cause of this misalignment by expanding transmission infrastructure would be ideal, but it would take years, if not decades~\cite{Chojkiewicz2024Transmission,Shawhan2025CleanPowerDelayed}. In this paper, we propose an algorithmic framework that leverages the flexibility of the transmission system to achieve significant savings using existing infrastructure. More precisely, by judiciously switching off a small number of transmission lines, we can alleviate congestion and reduce the cost and price of electricity throughout the system. This joint optimization of network topology and power generation is known as the Optimal Transmission Switching (OTS) problem~\cite{fisher2008optimal,fuller2012fast}, but its high computational cost has limited its adoption in routine operations.
Building on a recently developed foundation model for power system operations and insights from power engineering~\cite{yang2026gridsfm,bhan2026gridsfmfoundationmodelsolving}, we present an ML-guided framework that addresses these computational bottlenecks.

The need for faster and more reliable algorithms has always been a key impetus for power system research and engineering. System operators routinely solve problems such as unit commitment, economic dispatch and optimal power flow that determine how electricity should be generated and transmitted through large-scale grids~\cite{glover2012power}. OTS makes the optimization problem substantially harder by introducing the network topology itself as a decision: line-switching configurations create a combinatorial search, while each configuration must satisfy the nonconvex alternating current (AC) power flow equations. In recent years, advances in artificial intelligence (AI) and machine learning (ML) have created new opportunities to address this computational bottleneck, as learned models can exploit shared patterns across related optimization problem instances to guide algorithmic decisions ~\cite{huang2024distributional,scavuzzo2024machine,bengio2021machine}.


However, state-of-the-art ML models are not, by themselves, reliable optimizers for complex decision-making problems in power systems. Operating points in power grids must satisfy physical laws and engineering constraints, and in general there are no hard guarantees on the reliability of ML predictions in realistic operating scenarios. Therefore, the main body of work in power systems uses ML models in conjunction with classical methods~\cite{huang2027machine,bugaje2023realtime,han2023learning,divito2026diffusion,baker2020emulating}, aiming to achieve the speedup of ML while keeping the intuition and guarantees of existing methods, instead of fully replacing optimization with ML. 

In this paper, we apply this general approach to the AC-OTS problem. OTS in AC admits a Mixed-Integer Nonlinear Programming (MINLP) formulation and can, in principle, be solved with general-purpose MINLP solvers. Direct MINLP and relaxation-based approaches have been successful on small- and medium-scale benchmark systems~\cite{kocuk2017new,jain2024acopf}. However, AC-OTS combines two difficult sources of complexity: the continuous AC power-flow problem is nonlinear and nonconvex, while the line-status decisions introduce a combinatorial search over network topologies. This combination makes scaling to realistic transmission networks challenging. The need for fast solution methods is further amplified by increasingly dynamic operating conditions. Large computing loads, in particular, can exhibit substantial demand changes over short timescales~\cite{chen2025electricity,zhu2026evolving}. Recent ML-based approaches have also been shown to accelerate switching decisions in small- and medium-scale networks \cite{han2023learning,divito2026diffusion,bugaje2023realtime}. However, scaling these methods to larger systems might still incur substantial offline computational costs. In particular, training AC-OTS-specific models may itself require repeated expensive AC-OTS solves, for example through large numbers of episodes in reinforcement learning \cite{han2023learning} or in the generation of AC-OTS solutions for supervised learning~\cite{divito2026diffusion,bugaje2023realtime}. Thus, although learning can reduce online solution time, the computational burden of building the ML model can become a bottleneck as network size increases.

These limitations motivate the two design principles of our ML-guided AC-OTS framework. First, the framework should be able to adapt and scale to large networks without requiring expensive AC-OTS-specific training from scratch. Second, the learned model should provide useful signals for navigating the combinatorial search, while the feasibility and the operating cost should be verified with exact solvers. This separation makes the framework robust to imperfect learned predictions without relying on them for correctness.

Guided by these principles, our framework consists of three components. First, we construct a grid specific \emph{switching value model}, \model, by adapting GridSFM~\cite{yang2026gridsfm,bhan2026gridsfmfoundationmodelsolving}, a pretrained AC-OPF surrogate, to topology perturbations of each target network. Because fixing the switched lines reduces AC-OTS to AC-OPF on the resulting topology, GridSFM provides a natural foundation for estimating which switching actions are most effective in terms of generation cost. Second, we use insights from existing OTS heuristics to restrict the search to a set of candidate lines that are most promising for reducing operating costs. Third, we embed \model within a learning-guided search that ranks switching moves over this candidate set and selectively evaluates the most promising switching actions with an AC-OPF solver. In this way, learning guides where to search, while AC-OPF solvers verify the feasibility and true operating costs of the resulting switching actions. We develop three variants of this approach, Fixed-Rank Search, Adaptive-Rank Search, and Swap-Enhanced Search, which differ in how they incorporate the surrogate predictions to explore the topology space, offering different tradeoffs between search flexibility and computational cost.


Using existing benchmarks and realistic data center build-out scenarios in Texas, we demonstrate that our proposed methods can efficiently reduce the cost by optimizing line switching and operation decisions of existing transmission infrastructure within a tight computational budget of 50 seconds. In particular, for the 2,000-bus Texas network with approximately 9~GW of added data center load, our best algorithm reduces generation cost by 31.97\% and total load payments by 94.34\% by switching only five transmission lines. These results demonstrate that even an imperfect ML surrogate can effectively guide search in a challenging nonlinear combinatorial space, bringing AC transmission switching closer to real-time operational use. All code, case-construction scripts, data and baselines will be released upon acceptance.



More broadly, AC-OTS is representative of a wider class of optimization problems in power systems in which discrete decisions are coupled with expensive nonlinear AC network optimization. The same structure appears in problems involving contingency analysis \cite{hijazi2015optimal} and AC expansion planning \cite{das2019security}. Our results suggest a general paradigm for tackling such problems: use ML to capture useful structure in the underlying optimization landscape and to provide signals about where to search, rather than to replace optimization itself. Exact optimization methods can then retain responsibility for feasibility and solution quality.  This framework could provide a path towards solving increasingly complex grid-control and planning problems at operationally relevant scales.

\section{Background}
We begin by motivating transmission switching as a mechanism for reducing generation costs. We then formulate the AC optimal transmission switching (AC-OTS) problem. We also review the standard AC optimal power flow (AC-OPF) problem, a key building block of AC-OTS providing the underlying power-flow optimization model when the transmission switching decisions are fixed.

\subsection{Line Switching}
\begin{figure}[htbp]
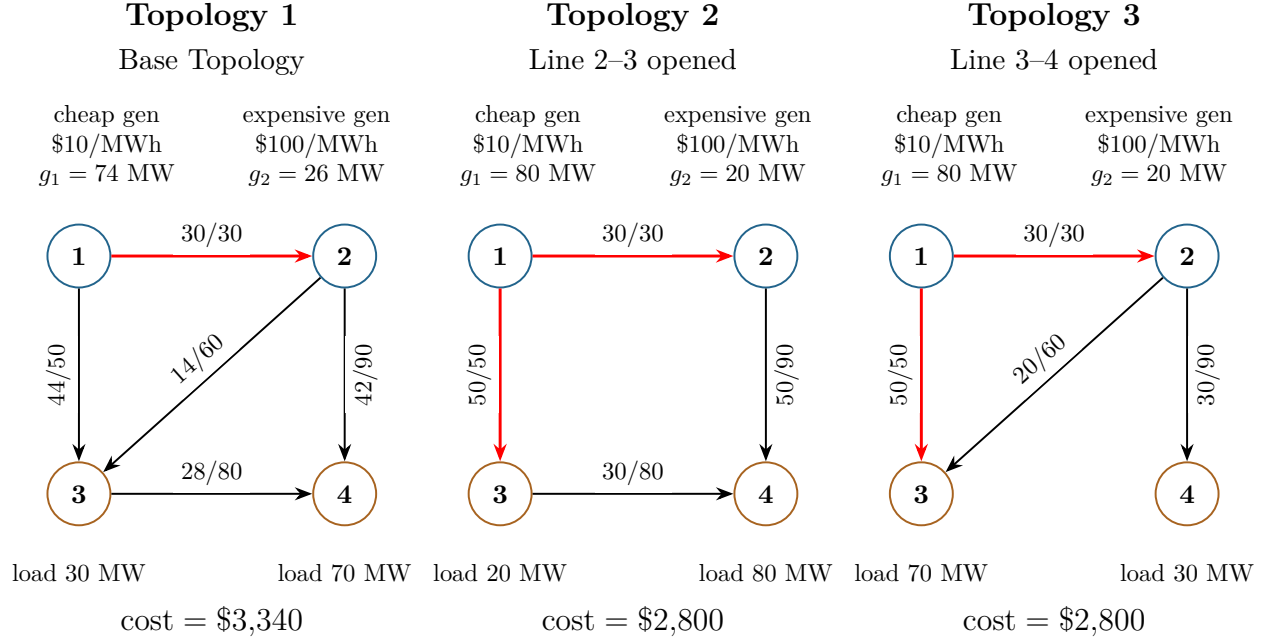

\centering
\networkdraw{1}{Base Topology}{74}{26}{30}{70}{3{,}340}\hfill
\networkdraw{2}{Line 2--3 opened}{80}{20}{20}{80}{2{,}800}\hfill
\networkdraw{3}{Line 3--4 opened}{80}{20}{70}{30}{2{,}800}
\caption{This example shows that switching off different lines is optimal for different load profiles. In all three systems, the total load is 100 MW. However, different topology would optimize generation cost for different bus-level load profiles. In particular, for the middle and right networks, the base topology is not optimal. Rather, disconnecting lines changes the congestion patterns, and lead to lower costs.}
\label{fig:concept}
\end{figure}

A major part of electricity prices is transmission congestion. When a transmission line reaches its thermal limit, generation must be redispatched to respect the binding network constraint, potentially requiring more expensive generators and creating spatial differences in prices. Switching off a subset of transmission lines can redistribute flows around such constraints and restore access to lower-cost generation  \cite{fisher2008optimal,hedman2011review}. This effect is counterintuitive because removing a transmission asset can improve, rather than degrade, the economic operation of a grid. Figure~\ref{fig:concept} illustrates this phenomenon: opening a transmission line changes the distribution of network flows and can yield a lower-cost dispatch than the original all-lines-in-service topology.

\subsection{AC Optimal Power Flow}

For a fixed network topology and load profile, the \emph{AC-OPF problem} determines how much active and reactive power each generator should produce, together with the voltage at each bus, so as to minimize generation cost while satisfying the nonlinear AC power-flow equations and operational constraints. This problem is nonconvex and has been a central object of study in power system operations~\cite{Low2026,capitanescu2016critical,coffrin2015qc}. Much progress has been made on solving it efficiently, but the problem remains difficult for large systems with thousands of buses, and solving one problem fully could take tens of seconds to minutes~\cite{cain2012history,lavaei2011zero,roald2023power}. Since AC-OPF is a basic building block of many other decision making tasks in power system operations, algorithms that provide meaningful speedups would advance a host of problems. 

We review the standard polar-coordinate formulation of the AC-OPF problem~\cite{glover2012power,Low2026}. We represent the transmission network by a set of buses $\mathcal{N}$ and a set of transmission lines $\mathcal{E}$, with $\mathcal{G}\subseteq\mathcal{N}$ denoting buses with controllable generation. Each bus $i\in\mathcal{N}$ has a fixed active and reactive demand, denoted by $p_i^{d}$ and $q_i^{d}$, respectively. The complex voltage at bus $i$ is represented in polar coordinates by its magnitude $v_i$ and phase angle $\theta_i$. At a generator bus $i\in\mathcal{G}$, $p_i^{\mathrm g}$ and $q_i^{\mathrm g}$ denote the active- and reactive-power outputs of the generator; for buses without controllable generation, these quantities are zero. 
A transmission line between bus $i$ and bus $k$ is modeled as a complex admittance $G_{ik}+j B_{ik}$, where $G_{ik}$ is called the conductance, $B_{ik}$ the susceptance, and $j$ the imaginary unit. We collect bus voltages, bus angles, generator active power and generator reactive powers into vectors, denoted by $\mathbf v$, $\boldsymbol{\theta}$, $\bm p$, and $\bm q$, respectively.  The AC-OPF minimizes generation cost subject to the power flow equation and engineering constraints:
\begin{subequations}\label{eq:acopf}
\begin{align}
\min_{\mathbf v, \boldsymbol{\theta},  \bm p, \bm q} \quad
& \sum_{i \in \mathcal{G}} c_{2,i} \, (p_i^{\mathrm g})^2 + c_{1,i} \, p_i^{\mathrm g} + c_{0,i}                   \label{eq:acopf-a}\\
\mathrm{s.t.}\quad
& p_i^{\mathrm g}-p_i^{\mathrm d}=v_i\sum_{k \sim i}v_k
 \bigl(G_{ik}\cos\theta_{ik}
      +B_{ik}\sin\theta_{ik}\bigr),
&& i\in\mathcal N,                                   \label{eq:acopf-b}\\
& q_i^{\mathrm g}-q_i^{\mathrm d}=v_i\sum_{k\sim i}v_k
 \bigl(G_{ik}\sin\theta_{ik}
      -B_{ik}\cos\theta_{ik}\bigr),
&& i\in\mathcal N,                                   \label{eq:acopf-c}\\
& \underline p_i^{\mathrm g}\le p_i^{\mathrm g}\le\overline p_i^{\mathrm g},
&& i\in\mathcal G,                                     \label{eq:acopf-d}\\
& \underline q_i^{\mathrm g}\le q_i^{\mathrm g}\le\overline q_i^{\mathrm g},
&& i\in\mathcal G,                                     \label{eq:acopf-e}\\
& \underline v_i\le v_i\le\overline v_i,
&& i\in\mathcal N,   \label{eq:acopf-f}\\
& \left[v_i v_k (G_{ik}\cos\theta_{ik}
      +B_{ik}\sin\theta_{ik})\right]^2 + \left[v_i v_k (G_{ik}\sin\theta_{ik}
      -B_{ik}\cos\theta_{ik})\right]^2 \le\overline S_\ell^2,
&& (i,k) \in\mathcal E,                                  \label{eq:acopf-g}
\end{align}
\end{subequations}
where we write $\theta_{ik} = \theta_i - \theta_k$ and $k \sim i$ when bus $k$ is connected to bus $i$. In the objective we adopt the commonly used quadratic cost functions, but other cost functions can be used as well. The active- and reactive-power balance constraints \eqref{eq:acopf-b} and \eqref{eq:acopf-c} ensure that, at each bus, local generation minus demand equals the net power injection determined by the AC power-flow equations. Thus, demand that cannot be met by local generation must be supplied through the network, with total generation covering both demand and transmission losses. The remaining constraints impose generator active- and reactive-power limits \eqref{eq:acopf-d} and \eqref{eq:acopf-e}, voltage magnitude limits~\eqref{eq:acopf-f}, and transmission-line flow limits~\eqref{eq:acopf-g}.

We can write~\eqref{eq:acopf-a} to \eqref{eq:acopf-g} in a more compact form as 
\begin{align} \label{eq:acopf-compact}
\min_{\mathbf{x}} \quad & \sum_{i \in \mathcal{G}} c_{2,i} (p_i^{\mathrm g})^2 + c_{1,i} \, p_i^{\mathrm g} + c_{0,i}  \qquad
\text{s.t.}\quad
\mathbf{x}\in \mathcal{F}_{\mathrm{AC}},
\end{align}
where the objective minimizes the generation cost over the variables $\mathbf{x}=(\bm{v}, \bm{\theta}, \bm p, \bm q)$ and $\mathcal{F}_{\mathrm{AC}}$ denotes the feasible set.

\subsection{AC Optimal Transmission Switching}

The \emph{AC optimal transmission switching (AC-OTS) problem} extends AC-OPF by jointly optimizing power-system operation and transmission topology.\cite{fisher2008optimal} It determines which transmission lines should be in service, in addition to deciding generator outputs and bus voltages, so as to minimize generation cost subject to the AC power-flow equations and operational constraints.\footnote{The actual operations of switching lines on and off are done electronically via automated breakers~\cite{IEEEC3704:2018}.} For notational simplicity, we formulate the problem in terms of switching lines off from a base topology, although our method applies equally when some lines can be switched on from a nominally off position.

Formally, the decision variables are \(S\) and \(\mathbf{x}\), where \(S \subseteq \mathcal{E}\) is the set of transmission lines to be opened and \(\mathbf{x}\) collects all the AC-OPF decision variables introduced above. We denote by \(\mathcal{F}_{\mathrm{AC}}(S)\) the set of AC-OPF solutions \(\mathbf{x}\) that are feasible for the network obtained by opening the lines in \(S\). The AC-OTS problem can then be written compactly as 
\begin{align} \label{eq:acots} \min_{S \subseteq \mathcal{E},\,\mathbf{x}} &\quad \sum_{i \in \mathcal{G}} c_{2,i} (p_i^{\mathrm g})^2 + c_{1,i} \, p_i^{\mathrm g} + c_{0,i}  \\ \text{s.t.} &\quad \mathbf{x} \in \mathcal{F}_{\mathrm{AC}}(S), \nonumber\\ &\quad |S| \leq J. \nonumber \end{align} 
Here, decisions are both the continuous AC power flow variables $\mathbf x$ and the discrete topological variables $S$. Hence, the generation costs are coupled with the selected transmission topology. This formulation includes the cardinality constraint \(|S|\leq J\) on the maximum number of lines that can be opened, which limits the extent to which the optimized topology can deviate from the base network. This constraint reflects the fact that AC-OPF only models the steady state behavior of the grid. Dynamic stability of the system is typically checked beforehand for the base topology and some of its close variants. Therefore, $J$ should not be too large. In addition, we show that most of the benefit in line switching is captured by a small number of switches.

We remark that AC-OTS is significantly more challenging to solve than the already difficult AC-OPF problem. On networks with thousands of buses and transmission lines, state-of-the-art optimization solvers struggle to find even an improving solution, despite the existence of topologies with substantially lower generation cost~\cite{holzer2025go,chevalier2026lowrank}.


\subsection{Prior Approaches to AC-OTS}


Since direct mixed-integer formulations of AC-OTS are not computationally tractable for medium- to large-scale grids, substantial work has focused on methods for finding high-quality switching decisions within practical time limits. These approaches broadly fall into three categories: approximations using DC power flow, AC-informed approximations, and learning-based methods that directly target the nonlinear AC-OTS problem.


The first class of approaches relies on DC approximations, replacing the nonlinear AC power-flow equations with a tractable linear model for identifying promising switching actions. Early work uses DC-OPF solutions to rank candidate lines~\cite{fuller2012fast}, while more recent methods leverage machine learning to prioritize candidate switches~\cite{yang2019line}, predict switching topologies~\cite{huang2027machine,bugaje2023realtime}, or learn topology decisions through differentiable DC-OPF layers~\cite{kim2025dispatchlearning}.
Although computationally attractive, these approaches inherit the limitations of the DC approximation. More precisely, switching decisions that appear promising may actually be infeasible for the full AC model~\cite{soroush2014accuracies}. Therefore, a list of candidates can often contain no good or even feasible actions for the AC-OTS problem. 

A second line of work seeks to retain the tractability of DC-OTS while incorporating information from the AC model. Feasibility can be improved by calibrating DC-model parameters using AC-OPF information~\cite{taheri2025ac}, or augmenting the DC-OTS formulation with AC admittance sensitivities and repair costs that approximate the effects of topology changes on AC operation~\cite{gupta2026acinformed}.

Finally, learning can also be applied directly to nonlinear topology optimization. A network- and objective-specific line-selection policy was trained with reinforcement learning~\cite{han2023learning}, and constraint-aware diffusion was used to predict complete switching configurations~\cite{divito2026diffusion}. These approaches, however, require substantial offline computation, often needing to generate tens of thousands of labels per network~\cite{han2023learning,divito2026diffusion}.   
This computational burden becomes increasingly significant as networks grow in size. The largest AC-OTS benchmark reported contains 500 buses~\cite{divito2026diffusion}, falling far short of the scale of large transmission networks of practical interest.


These limitations motivate methods that can search directly over AC-feasible topology changes without relying on expensive offline training.


\section{Methods}



 We develop a learning-guided system for real-time AC-OTS that uses fast, learned estimates to explore the combinatorial space of switching actions while reserving full AC-OPF solves (using conventional solvers) for evaluating promising candidate topologies. 
 
 Our system is guided by two design principles. The first is the ability to adapt and scale to large systems. 
 Rather than learning a network-specific switching policy from scratch, we provide a scalable recipe for constructing \model for each target grid. We begin with \textit{GridSFM}~\cite{yang2026gridsfm,bhan2026gridsfmfoundationmodelsolving}, a pretrained model for estimating AC-OPF solutions, and adapt it through a task-specific contrastive fine-tuning procedure. The resulting model is trained specifically to distinguish switching actions according to their effect on generation cost, rather than to perform generic cost prediction. 
The second is that the ML model and conventional solvers play distinct roles. The former provides guidance and the latter provide verification. We use \model to guide the combinatorial search and we do not require its estimates to guarantee feasibility or improvement in the true AC objective. Instead, \model must provide a sufficiently informative signal to identify promising moves, which are then evaluated using AC-OPF solves to determine their feasibility and true operating cost. This separation makes the system robust to imperfect learned estimates while retaining solver-based verification of every accepted topology. 
 
 Based on these principles, our system comprises three components. First, we construct a grid-specific \emph{switching-value model}, denoted by \model (Sec.~\ref{gridsfm_finetune}), that rapidly estimates the generation-cost reduction associated with opening transmission lines. Second, we use domain knowledge from existing OTS heuristics to identify a set of eligible lines that are most relevant to congestion relief  (Sec.~\ref{candidates}). Third, we embed \model within learning-guided local search algorithms that rank moves over the resulting neighborhoods and selectively evaluate the most promising candidates using an AC-OPF solver (Sec.~\ref{subsec:local search}). 

The three components jointly balance search breadth and computational cost in real-time AC-OTS. Although \model is substantially faster than AC-OPF, scoring every possible switching action remains too expensive within a real-time budget. Domain-informed filtering therefore narrows the search to lines most likely to affect congestion, keeping ML inference tractable. At the same time, \model allows us to retain a much broader set of candidate lines than would be practical with AC-OPF evaluations alone. Within this larger search space, learning-guided local search rapidly prioritizes promising moves and reserves expensive AC-OPF solves for selective verification. Together, these components enable broader yet computationally efficient exploration while ensuring that every accepted topology is verified for feasibility and operating cost.

\subsection{Grid-Specific Switching-Value Model}\label{gridsfm_finetune}

\begin{figure}[!htb]
\centering
\includegraphics[width=\linewidth]{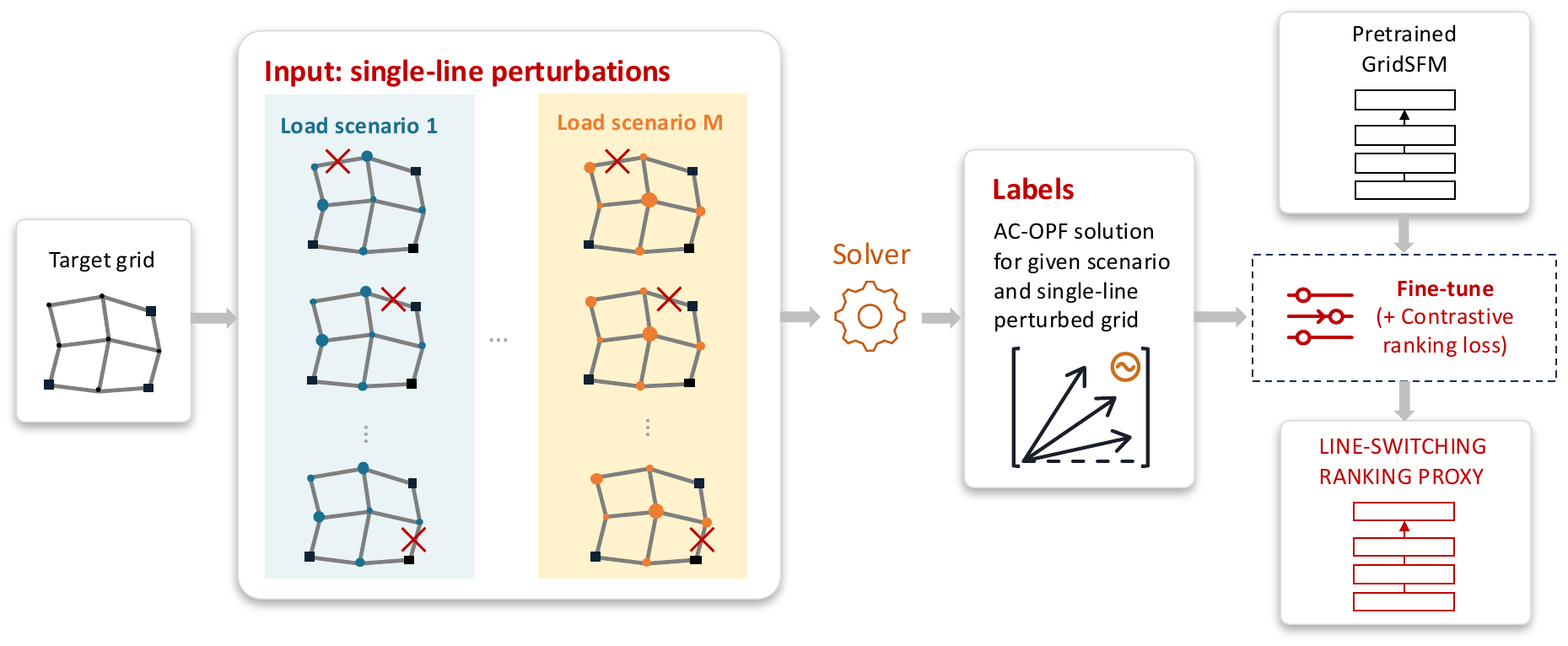}
\caption{Fine-tuning pipeline to build the \model. Training data combines load perturbations with single-line topology perturbations. Connectivity-preserving sub-networks are solved with exact solvers to obtain AC-OPF solutions for fine-tuning. Pretrained GridSFM is fine-tuned on sub-networks from each grid to obtain a grid-specific \model.}
\label{fig:pipeline}
\end{figure}


\subsubsection{GridSFM} \textit{GridSFM}~\cite{yang2026gridsfm,bhan2026gridsfmfoundationmodelsolving} is a graph transformer model that estimates solutions to the AC-OPF problem. The model has approximately 15 million parameters and operates directly on the structure of the transmission network. It represents a power system as a heterogeneous graph comprising buses, generators, loads, shunts, transmission lines, transformers, and network cycles, with edges encoding the physical relationships among these components. Network cycles---closed loops of transmission lines and transformers---are represented explicitly, allowing the model to capture coupling among line flows induced by Kirchhoff's voltage law. The model consists of blocks that combine linear attention among components of the same type with local message passing between physically connected components, thereby capturing interactions in the grid.
Its output heads predict AC-OPF decision variables, including generator active-power and voltage setpoints; a subsequent power-flow completion step recovers the remaining electrical state, from which generation cost can be evaluated. 

Importantly for our setting, GridSFM is pretrained jointly across a diverse collection of transmission networks and operating conditions rather than trained separately for a single grid. The model is therefore exposed during pretraining to variations in both operating conditions and network topology, allowing it to learn transferable relationships between network structure, power flow and generation cost. This is particularly relevant to AC-OTS, since once a set of lines $S\subseteq\mathcal{E}$ is opened, evaluating the resulting topology amounts to solving an AC-OPF problem on the modified network. GridSFM thus provides a natural pretrained starting point for rapidly assessing the effect of candidate topology changes.

GridSFM can be adapted to a target network using comparatively few solved AC-OPF instances, making it a natural foundation for constructing our grid-specific \model. Our objective, however, is not generic AC-OPF prediction. Instead, we require \model to discriminate among switching actions according to their effect on generation cost. We therefore introduce a task-specific contrastive fine-tuning procedure for this purpose, described next. We refer the reader to~\citet{bhan2026gridsfmfoundationmodelsolving} for complete details of the GridSFM architecture and training.

\subsubsection{Fine-tuning pipeline}

GridSFM is fine-tuned to capture how local topology changes affect the AC-OPF objective within a given grid. We create training data from both load perturbation and topology perturbation for each grid. For each target grid, we generate multiple load-perturbed operating scenarios. For each scenario, we construct sub-networks by removing one line, for all the switchable lines in the network, as illustrated in Figure~\ref{fig:pipeline}, discarding removals that disconnect the network. We solve AC-OPF on each remaining topology to obtain labels for fine-tuning.
We designed the data generation pipeline so that the number of labeled sub-networks grows linearly with the number of switchable lines in the network, for a fixed number of load scenarios. To keep offline data collection scalable, we avoid generating combinatorially many multi-line switching sub-networks, while still exposing \model to single-line changes that serve as the basic moves in our subsequent search procedure.



Internally, GridSFM represents each transmission line as an edge between its endpoint buses, and associate it with features such as admittance values and thermal ratings. \model augments the feature vector of a line by adding a binary \emph{in-service} indicator. This indicator is set to one for every line in the base topology, and switching off a line is represented by setting its indicator to zero. Thus, different candidate switching configurations can be represented by modifying the in-service indicators while retaining the same architecture and model parameters.


Next, we augment the physics- and cost-based losses used by GridSFM with a contrastive ranking objective that directly promotes promising switching actions.
For each load scenario, we rank the feasible single-line perturbations according to their AC-OPF generation cost and designate the $K$ lowest-cost candidates as positive examples, with the remaining candidates serving as negative examples. The resulting contrastive objective encourages \model to distinguish promising switching actions from less valuable ones. The value of $K$ controls how wide a band of promising topologies the model is encouraged to identify and is treated as a hyperparameter.

The contrastive objective emphasizes the \emph{relative ordering} of candidate topologies rather than requiring highly accurate absolute cost predictions. This distinction is important for the design of our search procedure: candidate topologies within the same operating scenario often have similar total costs, while the search uses only their relative ranking to decide which candidates should receive expensive AC-OPF verification. We therefore train the model to distinguish promising topology changes even when its absolute cost estimates are imperfect. 

Although the same framework could in principle use any ML model that estimated AC-OPF solutions, the ability to generalize across network topologies is particularly important for AC-OTS. Many existing AC-OPF prediction models are trained on a fixed network topology while varying only the operating conditions~\cite{pan2022deepopf,zhou2022deepopf}, whereas AC-OTS requires evaluating topologies that differ from the intact network. GridSFM is well suited to this setting because its pretraining spans transmission networks of different sizes and topologies, and its graph-based representation can naturally encode topology perturbations through the added binary \emph{in-service} line indicator. Fine-tuning on single-line removals then adapts the pretrained model to the local topology changes encountered on each target network during AC-OTS search.

\subsection{Candidate-Line Filtering}\label{candidates}

Although \model can score candidate switching actions much faster than AC-OPF, evaluating every possible line switching remains impractical for real-time decision making on large networks. For example, using a Nvidia H100 GPU, it would take GridSFM about 1 minute to evaluate the cost of switching every single line for a grid with thousands of buses. While significantly faster than conventional solvers, it would still take more than the allocated time budget in real-time operations.   We therefore first restrict the search to a set of candidate lines using two complementary filtering strategies from the OTS literature. The underlying observation is that topology impacts the cost through congestion.\footnote{Strictly, topology also affects cost through transmission losses. However, losses are typically less than a few percent in high-voltage AC networks, and opening a few lines changes them negligibly.} Thus, both strategies target lines that can affect congested parts of the network, either through topological proximity or through their estimated impact on congested flows.

Following the above intuition and existing OTS heuristics~\cite{fuller2012fast,sahraei2015real}, we create a congestion-region filter and retain lines in the vicinity of congested regions of the network. More precisely, we perform a breadth-first search extending two bus hops from both endpoints of each congested line, and every line incident to a visited bus is included in the candidate set. This filter captures nearby switching actions that may reroute power and change the congestion pattern~\cite{li2016real}. 

Topological proximity to congested lines alone can miss useful switching actions. Through Kirchhoff's voltage law, power flows are coupled along network cycles, which may span topologically distant parts of the grid. Consequently, opening a line far from a congested line can still substantially alter its flow. We account for these nonlocal effects using line-outage distribution factors (LODFs). These factors form a matrix, where the $(i,k)$'th element captures the sensitivity of the changes of power flows on line $i$ given a change on line $k$~\cite{guo2009direct}.
We create a LODF-based filter by adding the 50 lines outside of the congested region based on the largest LODF values. 

The resulting candidate set is deliberately much larger than those considered by existing OTS heuristics~\cite{li2016real,babaeinejadsarookolaee2023transmission}. We can afford this broader search space because we leverage the speed of \model for a fast estimation of the value of moves. This extended exploration increases the opportunity to identify lower-cost switching configurations.

\subsection{Learning-Guided Local Search}
\label{subsec:local search}
We now describe our learning-guided local-search framework, which combines the candidate-line filtering of Section~\ref{candidates} with the fast switching-value estimates provided by \model. We call a line \emph{eligible} if it is retained by the filtering step, and denote the set of eligible lines by $\mathcal E'$.
At any point in the search, the current network is represented by the set $S$ of opened lines, and a neighborhood consists of candidate moves that modify $S$ using only lines retained by the filtering step. Depending on the search variant, a move may open an additional line or simultaneously open a line and close a previously open one.

At each iteration, the local search alternates between two steps. \emph{Exploration.} Candidate moves are ranked according to their estimated generation-cost benefit using \model. Depending on the search variant, move values are either recomputed using the current topology by invoking \model or approximated using estimates obtained from the base topology, without further model inference. The search considers moves in order of estimated value, starting with the most promising. \emph{Verification.} A move is never accepted based on \model alone. Before accepting a candidate, we solve the full AC-OPF problem on the resulting topology to verify both its feasibility and actual generation cost. If the candidate is feasible and improves upon the incumbent, the move is accepted; otherwise, the search proceeds to the next candidate.

Once a move is accepted, the resulting topology becomes the new incumbent and the exploration--verification process repeats. The search continues until no further improving move is identified, the maximum number $J$ of opened lines is reached, or the prescribed time budget is exhausted.

We instantiate this learning-guided search in three variants that trade off the computational cost and quality of each search step, which we describe next; their pseudocodes are provided in the Appendix.

\subsubsection{Fixed-Rank Search (FRS)}\label{ranked_greedy}

\begin{figure}[!htb]
\centering
\includegraphics[width=\linewidth]{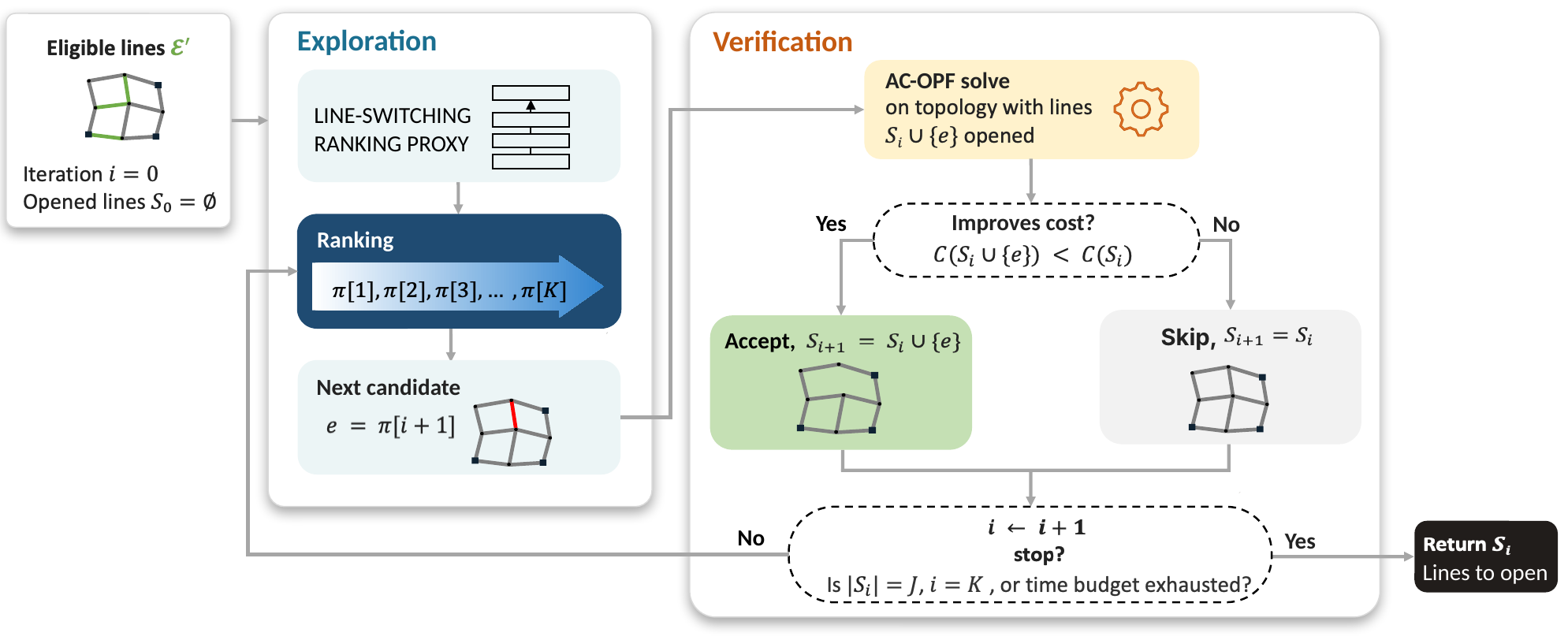}

\caption{Overview of the learning-guided search framework, illustrated for Fixed-Rank Search.
The eligible lines $\mathcal{E}'$ are first scored by the learned line-switching ranking model to define an exploration order $\pi$. At each iteration, the next candidate line is proposed and the resulting topology is verified using a full AC-OPF solve. The switching action is accepted only if it reduces the verified operating cost. This separation between learning-guided exploration and solver-based verification underlies all three proposed search variants.}
\label{fig:greedy}
\end{figure}

For any set of opened lines $S$, let $\widehat C_\theta(S)$ denote the generation cost predicted by \model for the topology obtained by opening the lines in $S$ from the base network. Fixed-Rank Search (FRS) is our least computationally expensive variant per search step: it invokes \model only once, before the search begins, and reuses the resulting ranking throughout the search. Its workflow is illustrated in Figure~\ref{fig:greedy}.

Specifically, FRS evaluates the removal of each eligible line $b\in\mathcal E'$ from the base topology and orders the lines by non-decreasing  predicted cost,
\[
\pi=(b_1,\ldots,b_{|\mathcal E'|}),
\qquad
\widehat C_\theta(\{b_1\})
\le \cdots \le
\widehat C_\theta(\{b_{|\mathcal E'|}\}).
\]
This ranking is computed once from the base topology and reused throughout the search, requiring no further \model inference as the topology evolves.

Starting from $S=\emptyset$, FRS considers lines sequentially according to $\pi$. For each line $b$, the candidate move is to open $b$, producing the candidate set of opened lines $S\cup\{b\}$. If the move preserves network connectivity, it undergoes the AC-OPF verification step described above. If the resulting topology is feasible and improves the incumbent cost, the move is accepted and $S\leftarrow S\cup\{b\}$; otherwise, $b$ is rejected and is not reconsidered in the future. This step is repeated until one of the stopping criteria described above is met.

\subsubsection{Adaptive-Rank Search (ARS)}

Adaptive-Rank Search (ARS) adapts its exploration step to the switching decisions of the previous iterations of the search. Unlike FRS, it does not reuse ranking information. Indeed, at every iteration, ARS invokes \model to reevaluate all eligible opening moves from the current set $S$ of opened lines.  Specifically, each eligible line $b\in\mathcal E'\setminus S$ defines the move $S\rightarrow S\cup\{b\}$, whose predicted cost is
\[
\widehat C_\theta(S\cup\{b\}).
\]
ARS ranks these moves by non-decreasing predicted cost and applies the AC-OPF verification step in this order until an improving feasible move is identified.

If the move associated with opening line $b$ is accepted, we update $S\leftarrow S\cup\{b\}$ and begin a new iteration, reevaluating all eligible opening moves from the updated set $S$. Note that, unlike FRS, an eligible line whose opening was rejected at an earlier iteration may become promising after other lines have been opened and is therefore reconsidered. This process is repeated until one of the stopping criteria described above is met.

\subsubsection{Swap-Enhanced Search (SES)}

Swap-Enhanced Search (SES) expands the local-search neighborhood by allowing previous switching decisions to be undone. In addition to an \emph{opening move} $S\rightarrow S\cup\{b\}$, SES considers \emph{swap moves} that re-close a line $a\in S$ while opening an eligible line $b\in\mathcal E'\setminus S$, producing the set of opened lines $(S\setminus\{a\})\cup\{b\}.$ 

SES has two phases. In the first phase, it computes the fixed ranking $\pi$ of FRS (Section~\ref{ranked_greedy}) and follows this search algorithm until five improving line openings have been accepted or the line-opening limit $J$ is reached. In the second-phase, it constructs two neighborhoods at each iteration. The \emph{opening neighborhood} consists of moves that open one of the ten highest-ranked eligible lines in the fixed FRS ranking $\pi$ that remain closed. The \emph{swap neighborhood} is formed by pairing each of these lines with each of the five most recently opened lines, yielding up to $5 \times 10 = 50$. SES invokes \model to evaluate every move in these neighborhoods.

The two moves with lowest predicted cost from each neighborhood are selected for AC-OPF verification. If at least one of these moves is feasible and improves upon the incumbent, the move with lowest verified cost is accepted and $S$ is updated accordingly. The neighborhoods are then rebuilt around the updated set of opened lines, and \model reevaluates all moves in the new neighborhoods.

\section{Evaluation}
\subsection{Networks}
We evaluate the proposed learning-guided methods on three transmission networks spanning a wide range of system sizes: the \texttt{api} variants of the IEEE 118-bus and Polish 2{,}737-bus networks from the PGLib-OPF benchmark library\cite{babaeinejadsarookolaee2019power}, and Texas2k, the 2{,}000-bus synthetic Texas grid (ACTIVSg2000, 2025-series summer peak)\cite{birchfield2017grid,xu2017application}. The \texttt{api} variants of the IEEE 118-bus and Polish systems are stressed operating cases with elevated loading and are commonly used to evaluate AC-OPF and AC-OTS algorithms under congested conditions. We use these two networks to evaluate the performance of our proposed methods on established benchmarks. Texas2k does not have an analogous \texttt{api} variant. To construct a comparably stressed setting, we therefore introduce data center loads at locations corresponding to publicly announced projects. For Texas2k, we consider the following scenarios, and 
additional details on the construction of the Texas2k scenarios are provided in the Appendix. In the networks studied, every line whose removal does not island the network is considered switchable.

\paragraph{Texas2k--data centers.} We construct a concentrated-load-growth scenario using publicly announced data center projects for which both location and planned capacity are available~\cite{crusoe2026abilene,dcd2025stargate,ecpkkr2025bosque,stream2024sanantonio}. The projects are mapped to geographically corresponding locations in Texas2k, adding approximately 9.1~GW of load in total, corresponding to a 10.7\% increase relative to the original 85.8~GW summer-peak demand. 

\paragraph{Texas2k--Fort Worth halved.} Because announced data center capacity is uncertain and may not fully materialize, we additionally consider a lower-load realization in the Fort Worth hub. We reduce the added Fort Worth data center load to 50\% of its value in the Texas2k--data centers scenario while leaving the added load at all other data center hubs unchanged.

\subsection{Baselines}
We compare our methods with six baselines: Price-difference sequential heuristic \cite{fuller2012fast} (PSeq), Random Congestion (RC), Congestion Simulated Annealing (CSA), AC-OTS with Gurobi, and DC-OTS with Gurobi, and Juniper \cite{kroger2018juniper}.

\paragraph{Price-difference sequential heuristic (PSeq).} This is the sequential LP heuristic of \citet{fuller2012fast}, adapted to the AC setting. Every in-service, non-islanding line $k$ is scored by the ranking parameter $\alpha_k = (\pi_{to} - \pi_{from}) \times P_k$, the price difference across the line times its active flow. A negative value marks counter-economic flow, from a higher-price bus to a lower-price one. We walk down the lines in order from the most negative $\alpha_k$. The first removal that lowers the AC-OPF cost is accepted.
\paragraph{Random Congestion (RC).} RC uses the same two-hop congestion and top cycle set as in the GridSFM-guided methods. Lines in these two sets are listed in random order, and this order remains fixed throughout the execution of the method. RC proceeds in this order and accepts a line to open if opening it lowers the AC-OPF cost.
\paragraph{Congestion Simulated Annealing (CSA).} CSA also uses the same two-hop congestion and top cycle set as in the GridSFM-guided methods. At each step it proposes a uniformly random move: open a line, reclose an opened line, or swap one for the other. Openings, including a swap's opening, come from the same candidate pool as random order, while reclosures are unrestricted. Each proposal is solved exactly and accepted if it lowers the AC-OPF cost.
\paragraph{Gurobi AC-OTS.} The exact optimal transmission switching problem in the polar AC formulation, built with PowerModels \cite{coffrin2018powermodels}. 
\paragraph{Gurobi DC-OTS.} The same switching problem in the linear DC power-flow approximation, for which the solution may not be AC-feasible.
\paragraph{Juniper.} The same AC optimal transmission switching model, solved by Juniper \cite{kroger2018juniper}, an open-source nonlinear branch-and-bound solver. Juniper solves continuous relaxations with Ipopt and uses a feasibility pump with HiGHS to find integer-feasible solutions.

\subsection{Metrics} 

\paragraph{Generation cost savings (\%).} We compare the methods using the percentage reduction in generation cost relative to the intact topology. Let $S$ denote the set of opened transmission lines found by a method, and let $C(S)$ be the corresponding generation cost obtained from a full AC-OPF solve on the resulting topology. The generation-cost savings are defined as
\begin{equation}
\operatorname{Generation\;Savings}(S)
=
100 \cdot
\frac{C(\emptyset)-C(S)}
{C(\emptyset)},
\end{equation}
where $C(\emptyset)$ is the AC-OPF generation cost of the intact topology with all lines in service.


\paragraph{Load payment savings (\%).} For Texas2k, we additionally report the percentage reduction in total load payment, as the realistic data center scenarios make the resulting prices and load payments more interpretable. Specifically, every solution to AC-OTS also yields nodal prices from its dual variables\footnote{In practice, the prices would be produced by the dual from solving the DC-OPF problem. In these cases, the duals from the AC-OPF problem and the DC-OPF are close to each other, and we report the AC-OPF duals for simplicity.}. The nodal price, also known as the locational marginal price, in \$/MWh , is the dual variable of the active power balance constraint at bus $i$, denoted by $\pi_i$~\cite{kirschen2026fundamentals}. It represents the price of supplying an additional unit of electricity demand at that location, given the current operating conditions.
Let $p^{d}_i$ denote the load served at bus $i$ in MW, and $p^{g}_k$ the dispatch of generator $k$ located at bus $k$. The hourly settlements are then computed as
\begin{equation}
\text{Total load payment:} \; \Pi \;=\; \sum_{i} \pi_i\, p^{d}_i ,
\quad
\text{Total generator revenue:} \; R \;=\; \sum_{k\in G} \pi_k\, p^{g}_k. 
\label{eq:payment}
\end{equation}

We note that the market price $\pi_i$ represents only a part of the utility bill, the other being fixed costs related to infrastructure and upgrades. Nevertheless, the increase in the prices and load payments would eventually be passed on to the customers. We compare the methods using the percentage reduction in total load payment relative to the intact topology. Let $\Pi(S)$ denote the total load payment under the topology with opened-line set $S$. The load-payment savings are defined as
\begin{equation}
\operatorname{Payment\;Savings}(S)
=
100\cdot
\frac{\Pi(\emptyset)-\Pi(S)}
{\Pi(\emptyset)},
\end{equation}
where $\Pi(\emptyset)$ is the total load payment under the intact topology.


\subsection{Computational Setup}
The maximum number of lines that can be opened ($J$) is set to 1\% of the total number of lines for each network. All methods use a time limit of 50s. We use MadNLP\cite{shin2024accelerating}, a GPU-accelerated interior point algorithm-based solver to solve the AC-OPF problems. All GridSFM-guided methods are run on NVIDIA H100 GPUs. The Juniper, DC-OTS and AC-OTS baselines are run on AMD EPYC 7542 CPUs. The software versions are Gurobi v12.0.3, Juniper 0.9.4, Ipopt 1.16.0, and HiGHS 1.25.4.
For algorithms that are randomized (RC and CSA), we report the results with 10 random seeds.



\section{Results}

\begin{table}[t]
\centering
\scalebox{1.00}{
\begin{tabular}{lrrrr}
\toprule
& \multicolumn{2}{c}{IEEE 118-bus-api (+62.1 \% load)}
& \multicolumn{2}{c}{Polish 2,737-bus-api (+84.0 \% load)} \\
\cmidrule(lr){2-3} \cmidrule(lr){4-5}
Method
& Gen. savings (\%)
& \# Lines
& Gen. savings (\%)
& \# Lines \\
\midrule

\rowcolor[HTML]{FFFFC5}
\multicolumn{5}{l}{\textit{Ours (50s)}} \\

\quad FRS
& 2.59
& 1
& \textbf{2.66}
& 15 \\

\quad ARS
& 2.59
& 1
& 1.89
& 9 \\

\quad SES
& \textbf{4.74}
& 1
& 2.41
& 7 \\

\midrule

\rowcolor[HTML]{FFFFC5}
\multicolumn{5}{l}{\textit{Baselines (50s)}} \\

\quad PSeq
& \textbf{4.74}
& 1
& 2.10
& 8 \\

\quad RC
& 0.79 $\pm$ 1.09
& 1.0 $\pm$ 0.0
& 1.69 $\pm$ 0.41
& 20.4 $\pm$ 2.1 \\

\quad CSA
& 3.08 $\pm$ 1.32
& 1.0 $\pm$ 0.0
& 0.94 $\pm$ 0.52
& 17.6 $\pm$ 8.5 \\

\quad Juniper
& 0.00
& 0
& 0.00
& 0 \\

\quad Gurobi AC-OTS
& 0.00
& 0
& 0.00
& 0 \\

\quad Gurobi DC-OTS
& 0.00
& 1$^{\dagger}$
& 0.00
& 0 \\

\midrule\midrule

\rowcolor{gray!20}
\multicolumn{5}{l}{\textit{Reference (24h)}} \\

\quad Juniper
& 4.74 (opt)
& 1
& 5.18
& 26 \\

\bottomrule
\end{tabular}
}
\caption{
Generation-cost savings and number of opened lines
for IEEE 118-bus-api and Polish 2,737-bus-api at a 50 s search budget.
Compared to the nominal cases in PGLib, the api case increases the load
by 62.1 \% and 84.0 \% in the IEEE 118-bus and Polish 2,737-bus
networks, respectively.
Seeded baselines report mean $\pm$ standard deviation over 10 seeds.
$\dagger$: the topology obtained by Gurobi DC-OTS is not AC-feasible. 
}
\label{tab:ots_results_118_2737}
\end{table}

\begin{table}[t]
\centering
\scalebox{0.75}{
\begin{tabular}{lrrrrrr}
\toprule
& \multicolumn{3}{c}{Texas2k--data centers (+10.7 \% load)}
& \multicolumn{3}{c}{Texas2k--Fort Worth halved (+9.9 \% load)} \\
\cmidrule(lr){2-4} \cmidrule(lr){5-7}
Method
& Gen. savings (\%)
& Payment savings (\%)
& \# Lines
& Gen. savings (\%)
& Payment savings (\%)
& \# Lines \\
\midrule

\rowcolor[HTML]{FFFFC5}
\multicolumn{7}{l}{\textit{Ours (50s)}} \\

\quad FRS
& 31.94
& 94.31
& 3
& 18.14
& 88.27
& 4 \\

\quad ARS
& 31.94
& 94.31
& 3
& 18.14
& 88.27
& 4 \\

\quad SES
& \textbf{31.97}
& \textbf{94.34}
& 5
& \textbf{18.16}
& \textbf{88.34}
& 5 \\

\midrule

\rowcolor[HTML]{FFFFC5}
\multicolumn{7}{l}{\textit{Baselines (50s)}} \\

\quad PSeq
& 23.73
& 78.12
& 1
& 14.06
& 79.39
& 1 \\

\quad RC
& 5.30 $\pm$ 10.23
& 24.14 $\pm$ 32.11
& 1.2 $\pm$ 1.5
& 8.66 $\pm$ 6.43
& 41.16 $\pm$ 30.99
& 4.8 $\pm$ 3.0 \\

\quad CSA
& 0.51 $\pm$ 0.72
& 7.07 $\pm$ 9.73
& 0.9 $\pm$ 0.6
& 3.53 $\pm$ 5.41
& 15.13 $\pm$ 25.87
& 2.7 $\pm$ 3.2 \\

\quad Juniper
& 0.00
& 0.00
& 0
& 0.00
& 0.00
& 0 \\

\quad Gurobi AC-OTS
& 0.00
& 0.00
& 0
& 0.00
& 0.00
& 0 \\

\quad Gurobi DC-OTS
& 0.00
& 0.00
& 0
& 0.00
& 0.00
& 0 \\

\midrule\midrule

\rowcolor{gray!20} 
\multicolumn{7}{l}{\textit{Reference (24h)}} \\

\quad Juniper
& 0.00$^{\ddagger}$
& 0.00
& 0
& 0.00$^{\ddagger}$
& 0.00
& 0 \\

\bottomrule
\end{tabular}
}
\caption{
Generation-cost savings, load-payment savings, and number of opened lines
for the Texas2k--data centers scenario and the Texa2k--Fort Worth halved scenarios, with a 50 s time limit. Relative to the nominal summer-peak case (85,758.9 MW), the Texas2k--data centers scenario and the Texa2k--Fort Worth halved scenarios increases the load by 10.7\% and 9.9\%, respectively.
Seeded baselines report mean $\pm$ standard deviation over 10 seeds.
$\ddagger$: no integral incumbent within 24 h.
}
\label{tab:ots_results_tx2k}
\end{table}

\subsection*{Benchmarks}
The results for IEEE 118-bus-api and Polish 2,737-bus-api are shown in Table \ref{tab:ots_results_118_2737}. On IEEE 118-bus-api, the proposed learning-guided Swap-Enhanced Search (SES) finds the optimal solution in terms of generation cost savings. On Polish 2,737-bus-api, both learning-guided FRS and SES outperform all other baselines. The solution found in 50 seconds is close to 50\% of the upper bound on savings found by Juniper using a 24-hour run. 

The results for the two Texas2k scenarios are shown in Table \ref{tab:ots_results_tx2k}. The learning-guided methods outperform all other baselines in both scenarios. SES performs the best in both generation-cost savings and payment savings in both scenarios. 

\subsection*{Texas2k data-center discussion}
\begin{figure}[!htb]
\includegraphics[width=\linewidth]{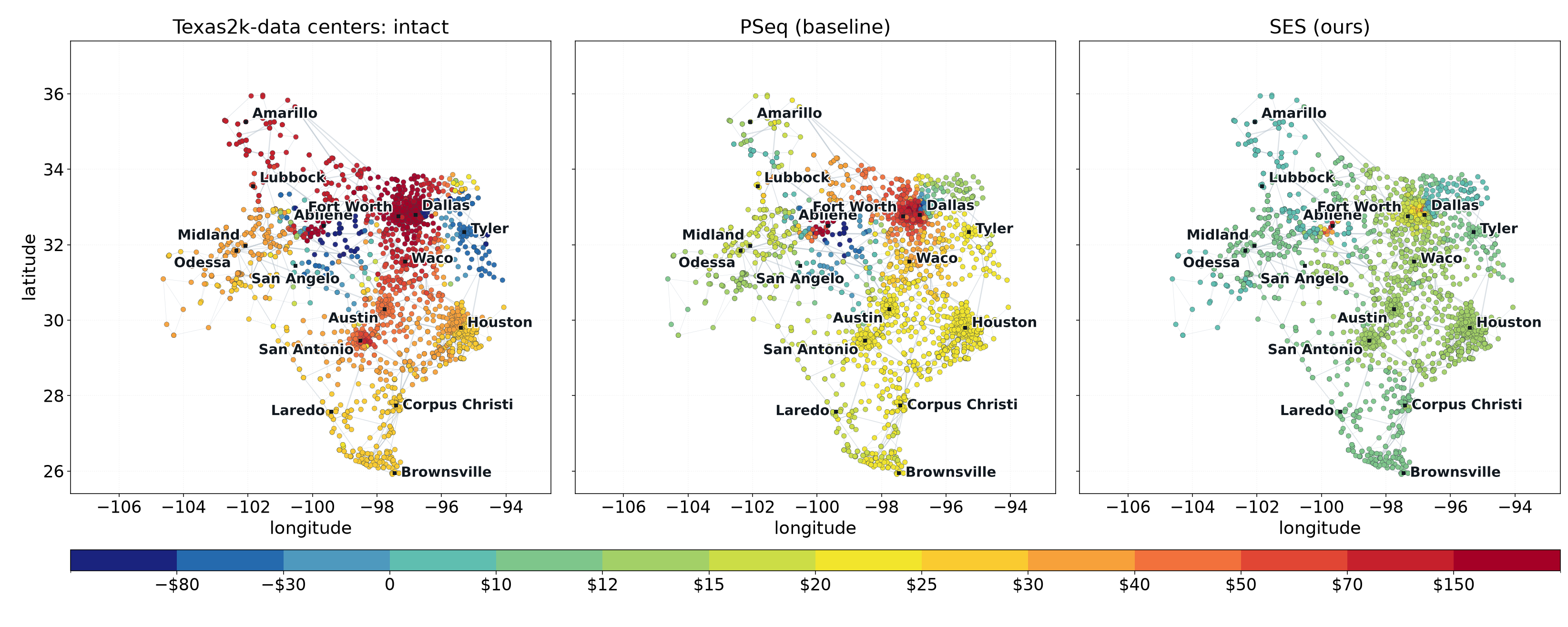}
\caption{Nodal prices on the Texas2k-data centers scenario: the intact network (left),
after switching with the PSeq baseline (1 line open, middle) and with SES (ours, 5 lines
open, right), each at 50\,s. Each dot is a bus colored by its nodal price. The color bands are shared by the three
panels and placed at quantiles of their pooled prices, rounded to readable values
(edges at \$$-80$, \$$-30$, 0, 10, 12, 15, 20, 25, 30, 40, 50, 70 and 150/MWh), so each
color covers a similar number of buses; negative prices are blue, positive prices run
from teal through yellow to dark red, and the color bar draws every band at equal
width. 
In the intact topology, 580 of the 2,751 buses exceed \$100/MWh (maximum \$13,316/MWh) and the
median price is \$32.91/MWh; with PSeq the median falls to \$22.14/MWh but 56 buses
around Dallas--Fort Worth and Abilene stay above \$100/MWh; with SES the median is
\$12.46/MWh and only 2 buses exceed \$100/MWh.
}
\label{fig:lmp}
\end{figure}
\begin{figure}[!htb]
\includegraphics[width=\linewidth]{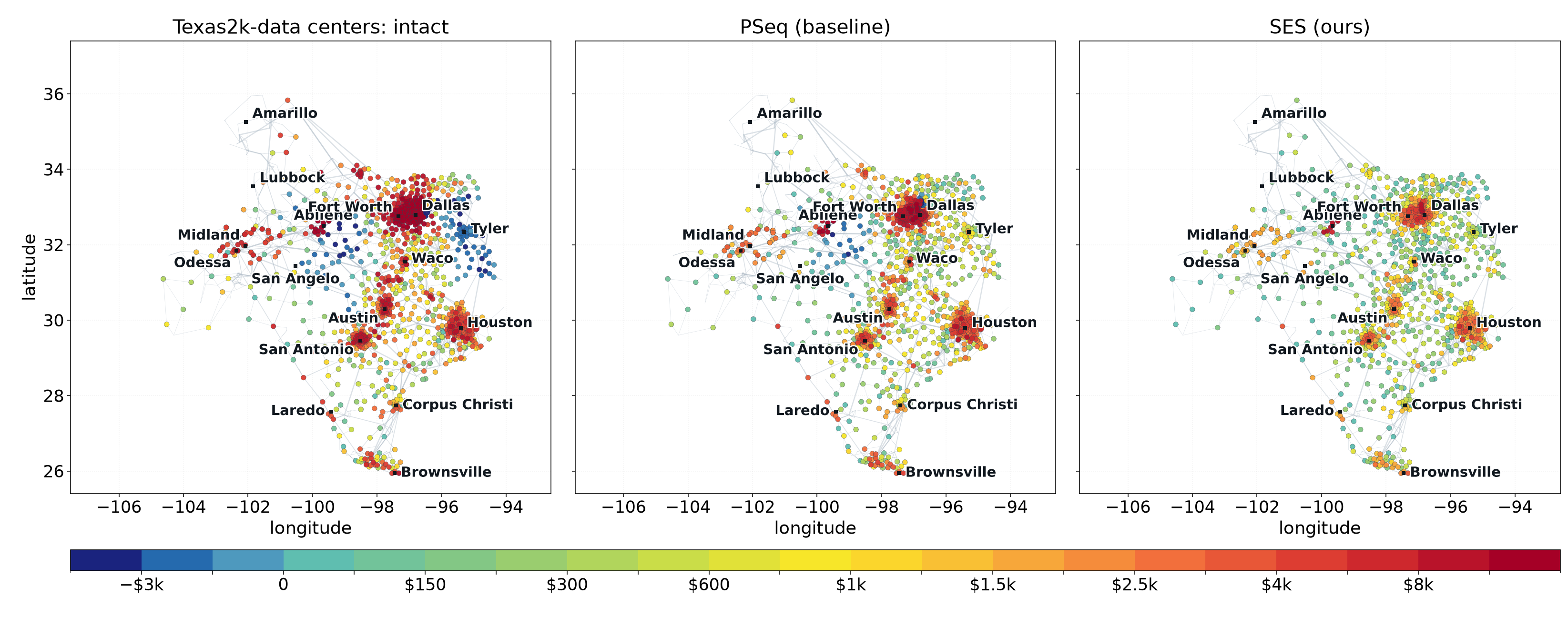}
\caption{Load payment per bus on the Texas2k-data centers scenario: the intact network
(left), after switching with the PSeq baseline (1 line open, middle) and with SES (ours,
5 lines open, right), each at 50\,s. Each dot is a bus with load, colored by its
load payment per hour, defined as the bus's nodal price multiplied by the load served at that bus; the total load payment is the sum over all
buses. Color bands are placed at quantiles of the pooled payments of the three panels. Total load payment is \$24.34M/h intact, \$5.33M/h with PSeq and \$1.38M/h with SES;
buses paying more than \$100k/h fall from 55 to 5 to none.
}
\label{fig:load}
\end{figure}


Under the Texas2k--data center scenario, the load payment is \$24.34M per hour for the base topology. Under the topology obtained with SES it pays \$1.38M per hour, a 94.34\% reduction. 
Figure~\ref{fig:lmp} shows nodal prices of the intact topology, topology found by the strongest baseline (PSeq), and topology found by our method (SES). With the intact topology, prices exceed \$100/MWh at 580 of the 2,751 buses, exceed \$1,000/MWh at 44 buses, and peaks at \$13,316/MWh. The area with higher nodal prices forms a corridor that runs from Amarillo through Lubbock and Abilene into
Dallas--Fort Worth and south toward Waco, with the most expensive buses concentrated around Abilene and Dallas--Fort Worth. With the five openings obtained by SES within 50\,s, the corridor with high prices clears: with only 2 buses remain above \$100/MWh, and the maximum drops to
\$169/MWh. The topology obtained by PSeq reduces the prices, but prices remain high at the Dallas--Fort Worth and Abilene areas. The corresponding load payment at each bus is shown in Figure \ref{fig:load}.


\begin{figure}[!htb]
\centering
\includegraphics[
    width=0.5\linewidth,
]{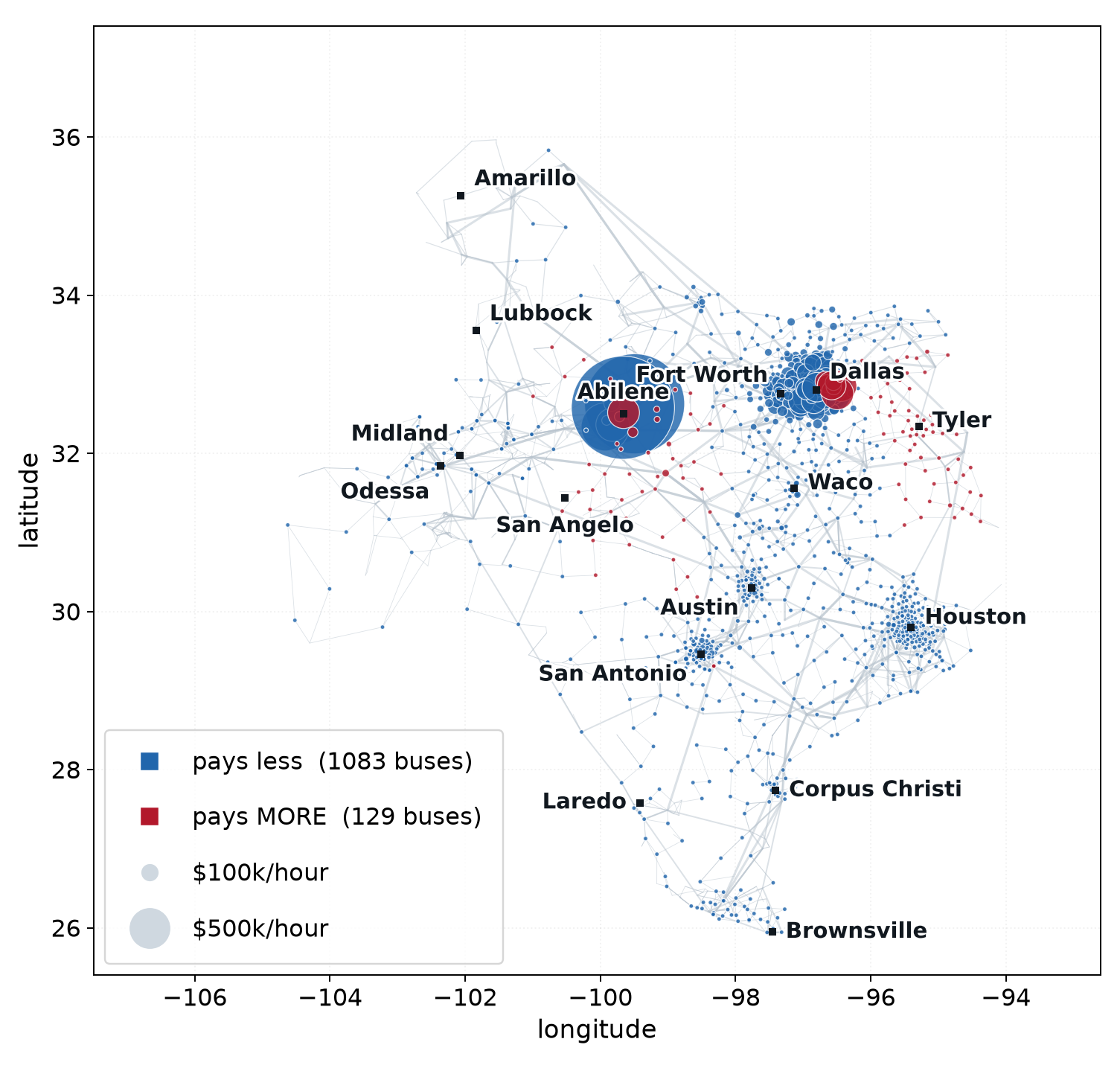}

\caption{
Change in load payment per bus after switching on Texas2k-data centers
(SES, 5 lines open, 50\,s time budget). Each circle is a bus with load; its area is proportional
to the change in what that load pays per hour (nodal price $\times$ load), relative to
the intact network. Blue: the bus pays less after switching (1,083 buses); red: it pays
more (129 buses). Total load payment falls from \$24.34M/h to \$1.38M/h.
}
\label{fig:change_load}
\end{figure}

Figure~\ref{fig:change_load} maps the change in load payment at each bus (blue: pays less; red: pays more; circle area proportional to the change in \$/h). In the data centers scenario, load pays less at 1,083 buses and more at
129. The reductions are
concentrated where nodal prices were highest: 94\,\% of the gross payment reduction
falls at the 311 buses whose intact price exceeded \$100/MWh, roughly 64\,\% of it in
Dallas--Fort Worth and 29\,\% around Abilene. The two largest single reductions, about \$2.9M/h and \$2.8M/h, are at Abilene. At the same time, a small number of buses pay modestly more (buses east of Dallas toward Tyler and some near the Abilene hub). The buses that pay more are almost all ones with a negative nodal price on the intact
network (119 of the 129).

\section*{Conclusion and Discussion}

We develop a novel learning-guided method to solve the AC Optimal Transmission Switching problem within real-time computational budgets. Our fine-tuning pipeline adapts a pretrained AC-OPF surrogate into a grid-specific switching-value model (\model) that estimates the impact of topology changes on generation cost. Domain-informed filtering keeps ML inference tractable, while fast ML-based evaluation enables exploration of a broader set of candidate lines than conventional methods. Within this search space, learning-guided local search prioritizes promising switching moves, which are subsequently verified using full AC-OPF solves to ensure feasibility and improvements in generation cost. Experiments on existing benchmarks and realistic load scenarios of the 2{,}000-bus Texas network show that the proposed ML-guided framework identifies switching decisions that significantly reduce the generation cost within a 50-second computational budget. Our results illustrate how ML and optimization can be combined to explore large-scale combinatorial search in power systems, offering a promising path toward tackling increasingly complex grid-control and planning problems at operationally relevant scales.

We also acknowledge several limitations. First, all benchmark networks considered in this study are synthetic test systems. Second, the search does not enforce (N-1) security of the resulting switched topologies~\cite{khanabadi2013optimal}. Third, while every accepted topology is verified to be AC-feasible, the proposed methods provide no guarantees of global optimality. Finally, we evaluate switched topologies as static operating points; transient stability, protection coordination, and the operational sequencing of multiple switching actions are outside the scope of the current model. Extending the framework to security-constrained switching, dynamic-security considerations, and multi-period operation is a natural direction for future work.

\section*{Appendix}
\subsection*{Texas2k case construction}\label{texas2kDC}
\subsubsection*{Texas2k--data centers}
Data center growth differs from conventional demand growth in both its spatial concentration and the speed at which large increments of load can appear. Individual facilities and campuses can reach hundreds of megawatts or more, potentially creating new transmission bottlenecks or substantially altering existing congestion patterns. Although Texas2k is a synthetic network, it is designed to approximate the structure of the ERCOT transmission system and provides geographic locations for buses and transmission lines~\cite{birchfield2017grid}. We therefore use it to construct a geographically grounded data center stress case based on publicly announced projects available as of September 2026.

At the time of writing, data center-related large-load interconnection requests in the Electric Reliability Council of Texas (ERCOT) service area exceed 420~GW~\cite{ercot2026queue}. This is several times larger than ERCOT's approximately 91~GW system peak demand, and the raw interconnection queue should therefore not be interpreted as a forecast of realized load. Rather than modeling the full queue, we use publicly announced projects for which both location and planned capacity are available~\cite{crusoe2026abilene,dcd2025stargate,ecpkkr2025bosque,stream2024sanantonio}.

We distribute the resulting build-out across ten geographic hubs according to the announced project pipeline where project-level information is available and according to established data center market shares elsewhere. This construction is intended to represent the geography of data center development rather than population. For example, Abilene receives the same 20\% share as Dallas despite its substantially smaller population because of its announced campus pipeline: approximately 1.2~GW has been developed, with a further 0.9~GW announced toward a total of 2.1~GW~\cite{crusoe2026abilene,dcd2025stargate}. The remaining hub shares, their project-level basis, and the source snapshot used to construct them are documented in the released case-construction script and accompanying README.

Within each hub, new demand is assigned to geographically corresponding Texas2k buses using the location labels supplied with the dataset. We use load-serving buses at 115 and 161~kV and any geographically matching bus at 230 and 500~kV. Individual injections are capped according to interconnection voltage, from 300~MW at 115~kV to 2,000~MW at 500~kV. These caps are scenario assumptions intended to prevent unrealistically large injections at lower-voltage transmission buses; very large campuses would in practice connect through higher-voltage infrastructure rather than through a single low-voltage substation. Each campus is represented as a lumped active and reactive power injection at the transmission bus with a power factor of 0.98. Customer substations, step-down transformers, collector networks and behind-the-meter generation are not modelled. The underlying Texas2k generation fleet and transmission network are otherwise unchanged.

\begin{figure}[!htb]
\includegraphics[width=\linewidth]{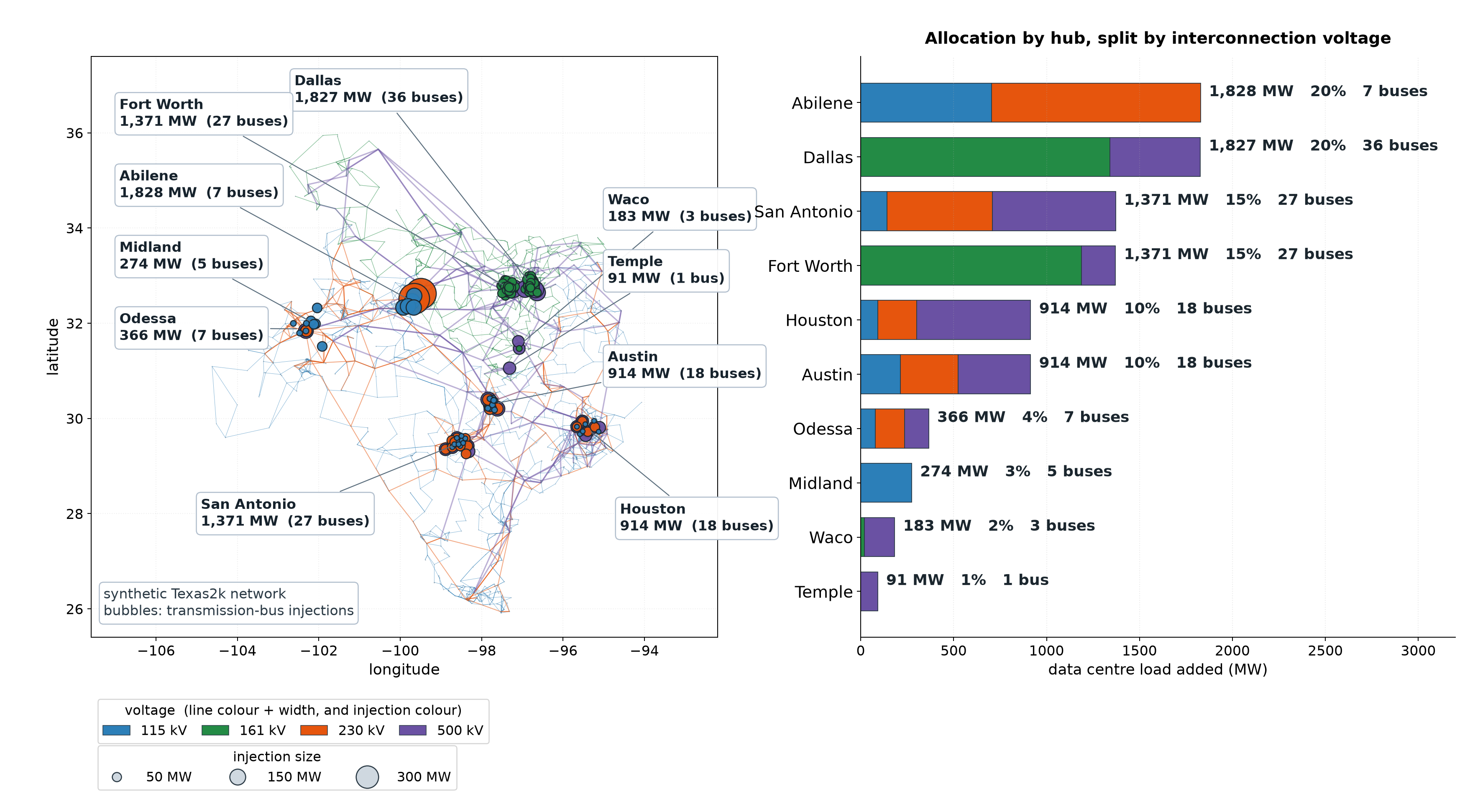}
\caption{\textbf{Constructed 9.1-GW data center stress scenario.} 149 synthetic load injections across ten metro clusters, colored by interconnection voltage and sized by MW, with the allocation by hub. The allocation follows announced projects and market weight, not population; individual points are eligible buses of the synthetic grid, not announced project locations.}
\label{fig:dcmap}
\end{figure}

We scale the total added data center demand by bisection to identify a highly stressed but AC-feasible operating point. Starting from the Texas2k 2025 summer-peak case, the total data center build-out is increased until the intact-network AC-OPF no longer returns a locally converged feasible operating point. The largest converged build-out is 9,139.5~MW, corresponding to a 10.7\% increase relative to the original 85.8~GW summer-peak demand. We use this point as the \textbf{Texas2k--data centers} scenario. The resulting total active demand is 94,898.4~MW, and the geographic distribution of the added load is shown in Figure~\ref{fig:dcmap}.

\subsubsection*{Texas2k--Fort Worth halved}
The Texas2k--data centers scenario represents a single highly stressed operating point near the network's estimated hosting-capacity boundary. We therefore evaluate sensitivity to uncertainty in the magnitude of geographically concentrated demand. The Dallas--Fort Worth metroplex is the dominant data center market in Texas and one of the largest in the United States~\cite{cbre2026datacenter}, but announced data center capacity is highly uncertain and may not fully materialize.
To represent a lower realization of the announced Fort Worth build-out, we construct the \textbf{Texas2k--Fort Worth halved} scenario by reducing all added data center load assigned to the Fort Worth hub to 50\% of its value in the Texas2k--data centers scenario. Added demand at all other hubs is left unchanged. This perturbation isolates the sensitivity of the switching solutions to uncertainty in one of the largest geographically concentrated load centers.

\subsection*{Pseudocode for GridSFM-guided methods}

\paragraph{Notation.}
Let $\mathcal{E}'\subseteq\mathcal{E}$ denote the set of candidate lines described in Section \ref{candidates}, excluding lines whose removal islands the intact network. Let $S\subseteq\mathcal{E}'$ denote the set of currently opened lines, $J$ the maximum number of simultaneously opened lines, and $T$ the wall-clock time budget. Given the current open set $S$, let $\mathcal I(S)$ denote the set of additional lines whose removal would island the network.

We use $\widehat C_\theta(S)$ to denote the GridSFM-predicted AC-OPF cost of the topology with lines $S$ open. Predictions for multiple candidate topologies are evaluated in a single batch. We use $C(S)$ to denote the cost returned by an exact AC-OPF solve and set $C(S)=+\infty$ if the solve is infeasible or fails to converge. The current best verified AC-OPF cost found so far is denoted by $c^\star$. A candidate topology is accepted only if $C(S)<c^\star$.

The pseudocode for \texttt{Fixed-Rank Search} and \texttt{Adaptive-Rank Search} are shown in Algorithm \ref{alg:greedy} and Algorithm \ref{alg:rerank}, respectively. \texttt{Swap-Enhanced Search} is shown in Algorithm \ref{alg:swap}, and it includes a \textsc{Fallback} procedure. In \texttt{Swap-Enhanced Search}, if none of the shortlisted moves improves the incumbent, \textsc{Fallback} scans the full fixed ranking \(\pi\), skipping already opened lines and previously solved or islanding topologies. When \(|S|<J\), it tests pure line openings; when \(|S|=J\), it tests swaps between each candidate and the \(A\) most recent openings in \(Q\). It returns the first topology with verified cost below \(c^\star\), or \(\emptyset\) if none is found before the ranking or time budget is exhausted.

\begin{algorithm}[H]
\caption{Fixed-Rank Search (\texttt{FRS})}
\label{alg:greedy}
\begin{algorithmic}[1]
\Require candidate set $\mathcal{E}'$, line limit $J$, time budget $T$
\State $S\gets\emptyset$, \quad $c^\star\gets C(\emptyset)$

\State $\pi \gets \textsc{Sort}\!\left(
    \mathcal{E}',\;
    b\mapsto \widehat C_\theta(\{b\})
\right)$
\Comment{ascending predicted cost; one-time ranking}

\For{$b$ in order $\pi$}
\If{$|S|=J$ \textbf{or} time $\ge T$}
\State \textbf{break}
\EndIf
\If{$b\in\mathcal I(S)$}
\State \textbf{continue}
\EndIf

\State $c\gets C(S\cup\{b\})$ \Comment{solver verification}
\If{$c<c^\star$}
    \State $S\gets S\cup\{b\}$, \quad $c^\star\gets c$
\EndIf

\EndFor

\State \Return $S$
\end{algorithmic}
\end{algorithm}

\begin{algorithm}[H]
\caption{Adaptive-Rank Search (\texttt{ARS})}
\label{alg:rerank}
\begin{algorithmic}[1]
\Require candidate set $\mathcal{E}'$, line limit $J$, time budget $T$
\State $S\gets\emptyset$, \quad $c^\star\gets C(\emptyset)$

\While{$|S|<J$ \textbf{and} time $<T$}
    \State $\pi \gets \textsc{Sort}\!\left(
        \mathcal{E}'\setminus\bigl(S\cup\mathcal I(S)\bigr),\;
        b\mapsto \widehat C_\theta(S\cup\{b\})
    \right)$
    \Comment{ascending predicted cost, rerank}

    \If{$\pi=\emptyset$}
        \State \textbf{break}
    \EndIf

    \State $\textit{improved}\gets\textbf{false}$
    \For{$b$ in order $\pi$}
        \If{time $\ge T$}
            \State \textbf{break}
        \EndIf

        \State $c\gets C(S\cup\{b\})$\Comment{solver verification}
        \If{$c<c^\star$}
            \State $S\gets S\cup\{b\}$, \quad $c^\star\gets c$
            \State $\textit{improved}\gets\textbf{true}$
            \State \textbf{break}
            \Comment{accept first improvement, then rerank}
        \EndIf
    \EndFor

    \If{\textbf{not} $\textit{improved}$}
    \State \textbf{break}
    \Comment{no improving line removal found}
\EndIf
\EndWhile

\State \Return $S$
\end{algorithmic}
\end{algorithm}

\begin{algorithm}[H]
\caption{Swap-Enhanced Search (\texttt{SES})}
\label{alg:swap}
\begin{algorithmic}[1]
\Require candidate set $\mathcal E'$, line limit $J$, time budget $T$;
         $A_{\max}=5$, $M=10$, $L=2$
\State $A\gets\min(A_{\max},J)$
\State $S\gets\emptyset$, \quad $c^\star\gets C(\emptyset)$
\State $\pi\gets\textsc{Sort}\!\left(
    \mathcal E',\, b\mapsto\widehat C_\theta(\{b\})
\right)$
\Comment{fixed ranking}

\Statex \textit{Phase 1: greedy}
\For{$b$ in order $\pi$}
    \If{$|S|=A$ \textbf{or} time $\ge T$} \textbf{break} \EndIf
    \If{$b\in\mathcal I(S)$} \textbf{continue} \EndIf
    \State $c\gets C(S\cup\{b\})$
    \If{$c<c^\star$}
        \State $S\gets S\cup\{b\}$, \quad $c^\star\gets c$
    \EndIf
\EndFor

\Statex \textit{Phase 2: open/swap local search}
\While{time $<T$}
    \State $B\gets$ first $M$ lines in $\pi$ not in $S$
    \State $Q\gets$ up to $A$ most recently opened lines in $S$

    \State $\mathcal N_{\rm open}\gets
    \{S\cup\{b\}:b\in B,\ |S|<J,\ b\notin\mathcal I(S)\}$

    \State $\mathcal N_{\rm swap}\gets
    \{(S\setminus\{a\})\cup\{b\}:
      a\in Q,\ b\in B,\ b\notin\mathcal I(S\setminus\{a\})\}$

    \State $\mathcal L\gets
    \textsc{Top}_{L}(\mathcal N_{\rm open},\widehat C_\theta)
    \cup
    \textsc{Top}_{L}(\mathcal N_{\rm swap},\widehat C_\theta)$
    \Comment{surrogate shortlist}

    \State exactly solve all previously unsolved $S'\in\mathcal L$

    \If{$\min_{S'\in\mathcal L} C(S')<c^\star$}
        \State $S\gets\arg\min_{S'\in\mathcal L} C(S')$,
               \quad $c^\star\gets C(S)$
    \Else
        \State $S'\gets\textsc{Fallback}(S,\pi,Q,J,T)$
        \Comment{first verified improvement}
        \If{$S'=\emptyset$} \State \textbf{break} \EndIf
        \State $S\gets S'$, \quad $c^\star\gets C(S')$
    \EndIf
\EndWhile

\State \Return $S$
\end{algorithmic}
\end{algorithm}

\section*{Data availability}
All two Texas2k evaluation cases, their generation receipts, the runs behind every table entry and figure, and complete per-grid recipes that regenerate the training data will be released with the code upon publication, with an archived private snapshot available to reviewers. Benchmark networks: PGLib-OPF (\url{https://github.com/power-grid-lib/pglib-opf}) and the Texas A\&M ACTIVSg2000/Texas2k repository (\url{https://electricgrids.engr.tamu.edu/electric-grid-test-cases/activsg2000/}).

\section*{Code availability}
The code for the search methods, the GridSFM fine-tuning recipe, the AC-OPF verification pipeline, all baselines and the case-construction scripts is available to reviewers through an archived private snapshot and will be publicly released upon publication.

\section*{Acknowledgments}
This work was done during Weimin Huang’s internship at Microsoft Research and at the University of Southern California. The research is partially supported by the National Science Foundation (NSF) grant \#2112533: “NSF Artificial Intelligence Research Institute for Advances in Optimization”, the NSF grant \#2346058: “NRT-AI: Integrating Artificial
Intelligence and Operations Research Technologies”.

\section*{Author contributions}
{\setlength{\parindent}{0pt}\setlength{\parskip}{4pt}

\textbf{Weimin Huang} developed the learning-guided search methods, conducted the experiments and co-wrote the manuscript.

\textbf{Weiwei Yang} adapted and fine-tuned GridSFM to construct the Line-switching Ranking Proxy, set up the initial experiments and datasets, and co-wrote the manuscript.

\textbf{Bistra Dilkina} provided research oversight, gave feedback on the optimization methodology and edited the manuscript.

\textbf{Konstantina Mellou} mentored Weimin Huang, advised on the design of the search methods and edited the manuscript.

\textbf{Ishai Menache} provided research oversight, gave feedback on the optimization methodology and edited the manuscript.

\textbf{Marco Molinaro} mentored Weimin Huang, advised on the design of the search methods, assisted with the experiments and edited the manuscript.

\textbf{Chris White} provided research oversight and edited the manuscript.

\textbf{Baosen Zhang} provided research oversight and power systems and AI expertise, and co-wrote the manuscript.

}

\section*{Competing interests}
The authors declare no competing interests. 

\section*{Correspondence}
Correspondence and requests for materials should be addressed to Weiwei Yang (\href{mailto:weiwei@asemic.ai}{weiwei@asemic.ai}).

\bibliographystyle{unsrtnat}
\bibliography{sample-base}

\end{document}